\pdfoutput=1

\documentclass[sigconf,screen]{acmart}

\DeclareMathDelimiter{(}{\mathopen} {operators}{"28}{largesymbols}{"00}
\DeclareMathDelimiter{)}{\mathclose}{operators}{"29}{largesymbols}{"01}

\usepackage[normalem]{ulem} 
\usepackage{ifthen}  
\usepackage{color} 

\usepackage[utf8]{inputenc} 
\usepackage{microtype} 

\usepackage{subcaption}

\graphicspath{{./figures/}}
\DeclareGraphicsExtensions{.pdf,.png}
\usepackage{tikz} 
\usepackage{float} 

\usepackage{vwcol}
\usepackage{multirow}  
\usepackage{array}  
\usepackage{varwidth} 
\newcolumntype{C}[1]{>{\centering\arraybackslash}p{#1}}

\usepackage{amsmath,amsthm,amsfonts}
\mathchardef\hyphenmathcode=\mathcode`\-
\newtheoremstyle{custom}
{2pt}
{2pt}
{}
{0pt}
{\slshape}
{}
{ }
{\thmname{#1}:\thmnote{ (#3)}}
\theoremstyle{custom}
\newtheorem{example}{Example}
\theoremstyle{definition}
\newtheorem{definition}{Definition}

\usepackage{stmaryrd}

\usepackage{pifont} 
\usepackage{xspace}  
\usepackage{cleveref}
\newcommand{\MP}[1]{\textcolor{red}{#1}}

\newboolean{publicversion}
\setboolean{publicversion}{true}

\ifthenelse{\boolean{publicversion}}{
\newcommand{\grumbler}[2]{}
\newcommand{\editing}[1]{}

}{
\newcommand{\grumbler}[2]{\MP{{\bf #1}: #2}}

\newcommand{\editing}[1]{{\color{red}#1}}

}

\newcommand{\al}[1]{\grumbler{\scriptsize OO}{\scriptsize \bf #1}}
\newcommand{\bk}[1]{\grumbler{\scriptsize BK}{\scriptsize \bf#1}}
\newcommand{\ms}[1]{\grumbler{\scriptsize MS}{\scriptsize \bf#1}}

\newcommand{\status}[3]{}

\usepackage{listings}
\lstdefinestyle{embedded}{
    language=Python,
    basicstyle=\ttfamily\small,
    numbers=left,
    numberstyle=\tiny,
    numbersep=3pt,                  
    numberblanklines=true,
    frame=tb,
    aboveskip=5pt,
    belowskip=5pt,
    columns=fullflexible,
    showstringspaces=false,
    keepspaces=true,
    showlines=true,
    xleftmargin=8pt,
    backgroundcolor=\color{gray!4},
    framesep=1pt,
}
\lstdefinestyle{asm}{
    language={[x86masm]Assembler},
    basicstyle=\ttfamily\small,
    numbers=left,
    numberstyle=\tiny,
    numbersep=3pt,                  
    numberblanklines=true,
    frame=tb,
    aboveskip=0pt,
    belowskip=0pt,
    columns=fullflexible,
    showstringspaces=false,
    keepspaces=true,
    showlines=true,
    xleftmargin=8pt,
    backgroundcolor=\color{gray!4},
    morekeywords={MFENCE,R14,RAX,RBX,RCX,RDX,CMOVNL,CMOVNZ},
    morecomment=[l]{\#},
    commentstyle=\color{gray},
    keywordstyle=\ttfamily,
}

\lstdefinelanguage{yaml}{
    morecomment=[l]{\#},
    commentstyle=\color{gray}
}

\let\origlstlisting=\lstlisting
\let\endoriglstlisting=\endlstlisting
\renewenvironment{lstlisting}
{\mathcode`\-=\hyphenmathcode
\everymath{}\mathsurround=0pt\origlstlisting}
{\endoriglstlisting}

\newcommand{\myparagraph}[1]{\smallskip \noindent{\textbf{{#1}.}}}
\newcommand{\myparagraphnodot}[1]{\smallskip \noindent{\textbf{{#1}}}}
\newcommand{\code}[1]{\mbox{\texttt{#1}}}
\newcommand{\circled}[1]{\raisebox{.5pt}{\textcircled{\raisebox{-.9pt} {#1}}}}

\newcounter{chal}
\newcommand{\challenge}[1]{\smallskip\par\refstepcounter{chal} \noindent{\textbf{CH\thechal: #1.}}}

\newcommand{\secref}[1]{$\S$\ref{sec:#1}}
\newcommand{\subsecref}[1]{$\S$\ref{subsec:#1}}

\newcommand{\figref}[1]{Figure~\ref{fig:#1}}

\newcommand{\tabref}[1]{Table~\ref{tab:#1}}

\newcommand{\defref}[1]{Def~\ref{def:#1}}

\newcommand{\chref}[1]{CH\ref{ch:#1}}

\newcommand{\sys}{Revizor}
\newcommand{\method}{MRT}
\newcommand{\methodlong}{Model-based Relational Testing}

\newcommand{\ctseq}{\textit{CT-SEQ}}
\newcommand{\ctcb}{\textit{CT-COND}}
\newcommand{\ctsbp}{\textit{CT-BPAS}}
\newcommand{\ctcbsbp}{\textit{CT-COND-BPAS}}

\newcommand{\archseq}{\textit{ARCH-SEQ}}

\newcommand{\memseq}{\textit{MEM-SEQ}}
\newcommand{\memcond}{\textit{MEM-COND}}

\newcommand{\seq}{\textit{SEQ}}
\newcommand{\cond}{\textit{COND}}
\newcommand{\bpas}{\textit{BPAS}}

\newcommand{\ibasic}{\texttt{AR}}
\newcommand{\ibm}{\texttt{AR+MEM}}
\newcommand{\ibc}{\texttt{AR+CB}}
\newcommand{\ibcm}{\texttt{AR+MEM+CB}}

\newcommand{\ibmv}{\texttt{AR+MEM+VAR}}
\newcommand{\ibcmv}{\texttt{AR+MEM+CB+VAR}}

\newcommand{\program}{Prog}
\newcommand{\archstate}{Data}
\newcommand{\microstate}{Ctx}
\newcommand{\htrace}{HTrace}
\newcommand{\ctrace}{CTrace}
\newcommand{\attack}{Attack}
\newcommand{\contract}{Contract}

\newcommand{\term}[1]{\textbf{#1}}

\setcopyright{none}
\renewcommand\footnotetextcopyrightpermission[1]{}
\acmConference[ ]{_}
\acmBooktitle{}
\begin{document}
    
  \title{\sys{}: Testing Black-Box CPUs against Speculation Contracts}

  \author{Oleksii Oleksenko}
  \authornote{Work partially done at Microsoft Research Cambridge.}
  \author{Christof Fetzer}
  \affiliation{%
    \institution{TU Dresden}
    \city{Dresden}
    \country{Germany}
  }

  \author{Boris Köpf}
  \affiliation{%
    \institution{Microsoft Research}
    \city{Cambridge}
    \country{UK}
  }

  \author{Mark Silberstein}
  \affiliation{%
    \institution{Technion}
    \city{Haifa}
    \country{Israel}
  }

  \renewcommand{\shortauthors}{}


\begin{abstract}

    Speculative vulnerabilities such as Spectre and Meltdown expose speculative execution state that can be exploited to leak information across security domains via side-channels.
    Such vulnerabilities often stay undetected for a long time as we lack the tools for systematic testing of CPUs to find them.

    In this paper, we propose an approach to \emph{automatically} detect microarchitectural information leakage in commercial black-box CPUs.
    We build on speculation contracts, which we employ to specify the permitted side effects of program execution on the CPU's microarchitectural state.
    We propose a \methodlong{} (\method{}) technique to empirically assess the CPU compliance with these specifications.

    We implement \method{} in a testing framework called \sys{}, and showcase its effectiveness on real Intel x86 CPUs.
    \sys{} automatically detects violations of a rich set of contracts, or indicates their absence.
    A highlight of our findings is that \sys{} managed to automatically surface Spectre, MDS, and LVI, as well as several previously unknown variants.
    \vspace{0.5cm}
\end{abstract}




  \maketitle
  \vspace{0.6cm}


\section{Introduction}
\label{sec:introduction}

The instruction set architecture (ISA) specifies the functional behavior of a CPU but abstracts from its implementation details (microarchitecture).
This abstraction enables rapid development of hardware optimizations without requiring changes to the software stack;
unfortunately, it also obscures the security impact of these optimizations.
Over the last decade researchers discovered numerous microarchitectural zero days, including Spectre-style attacks that use microarchitectural state to exfiltrate secret information obtained during transient execution~\cite{Kocher2018,Lipp2018}.
The problem is expected to get worse as Moore's law subsides and CPU architects are compelled to apply ever more aggressive optimizations~\cite{Rodrigo2021}.

Speculation contracts (short: contracts)~\cite{Guarnieri2021} have been proposed as a way out of this situation by providing a \emph{specification} of the microarchitectural side effects.
Contracts declare which ISA operations an attacker can observe through a side channel, and which operations can speculatively change the control/data flow.
For example, a contract may state: an attacker can observe addresses of memory stores and loads, and the CPU may mispredict the targets of conditional jumps.
If a CPU implementation permits the attacker to observe \emph{more} than that (e.g., addresses of loads after mispredicted \emph{indirect} jumps), the CPU violates the contract, indicating an unspecified leak in the microarchitecture.

For software developers, contracts are a foundation for microarchitecturally secure programming: they spell out the assumptions that are required for checking that mitigations are effective and code is free of leaks.
For example, a recent survey~\cite{cauligi2021sok} classifies existing tools for detecting speculative vulnerabilities in the language of contracts.
For hardware developers, contracts can provide a target specification that describes the permitted microarchitectural effects of the CPU's operations, without putting further constraints on the hardware implementation.
Thus, contracts hold the promise to achieve for speculative vulnerabilities what consistency models have provided for memory consistency~\cite{alglave2012formal}.

Despite the contracts' potential, so far they have only been used for establishing security guarantees of small white-box models of CPUs with toy ISAs~\cite{Guarnieri2021}.
In the context of real-world CPUs, several existing tools (e.g., Medusa~\cite{Moghimi2020a}, SpeechMiner~\cite{Xiao2020}, and CheckMate~\cite{Trippel2018}) target automated detection of known types of speculative leaks, but not contract violations in general.
Thus, it has been an open challenge to test contract compliance of real-world CPUs, with complex ISAs and absent (or intractable) models of the microarchitecture.

\myparagraph{Approach} 
In this paper, we propose a method and a tool for testing real-world CPUs against speculation contracts.
Our method, called \emph{Model-based Relational Testing} (MRT), is a randomized search for ``evidence'' of contract violations, i.e., for counterexamples to contract compliance.

Such a counterexample is a specific instance where the CPU leaks more information than the contract permits.
In particular, a counterexample is an instruction sequence together with a pair of inputs that produce the \emph{same} observations according to the contract (\emph{contract trace}), but \emph{different}  microarchitectural side-effects on the CPU (\emph{hardware trace}).

MRT searches for counterexamples by creating samples---random instruction sequences (\emph{test cases}) together with random inputs---and checking if any of them constitutes a counterexample.
A key observation is that this check \emph{does not} require an explicit model of the microarchitecture.
This is because one only needs to compare traces of the same kind, that is, contract traces to contract traces, and hardware traces to hardware traces.
This enables side-stepping the need for establishing a connection between them via a model of the microarchitecture (as done in~\cite{Guarnieri2021}) and enables testing commercial black-box CPUs.
However, the search for counterexamples on real-world CPUs poses a new set of challenges:

The first challenge is to cope with the \emph{intractable search space}:
ISAs typically include hundreds of instructions, dozens of registers, and permit large memory spaces.
This creates an intractable number of possibilities for both test cases and for inputs to them.
Moreover, there are no means to measure coverage for black-box CPUs, which precludes a guided search.
We solve this problem by using an incremental generation process that aims to create ample opportunities for speculation:
(1) We perform testing in rounds, where we start by generating short instruction sequences with few basic blocks, a small subset of registers, and where we confine all memory accesses to a narrow memory range.
(2) After each round without counterexample, we invoke a diversity analysis that counts the number of tested instruction patterns that we expect to induce speculative leaks.
This analysis triggers reconfiguration of the test generator to gradually expand the search space in subsequent testing rounds.

The second challenge is to obtain \emph{deterministic hardware traces} from modern high-performance CPUs with complex and unpredictable microarchitectures.
For this we (1) create a low-noise measurement environment where we execute test cases in complete isolation and perform a side-channel attack (e.g., Prime+Probe on the L1D cache) to detect leakage into the microarchitecture, and (2) we control the microarchitectural context using a technique we call \emph{priming}:
Priming collects traces for a large number of pseudorandom inputs to the same test case \emph{in sequence}.
In this way, execution with one input effectively sets the microarchitectural context for the next input.
This enables collection of hardware traces with predictors primed in a diverse but deterministic fashion, which is key to obtaining comprehensive and stable hardware traces.

The third challenge is to \emph{generate contract traces for complex ISAs} such as x86.
To tackle this challenge we implement executable contracts by instrumenting an existing ISA emulator with a checkpointing mechanism similar to~\cite{Oleksenko2020}, which enables us to explore correct and mispredicted execution paths, and to record the contract-prescribed observations during program execution.

\myparagraph{Tool \& Evaluation}
We implement MRF as a testing framework \sys{}\footnote{Revizor is a name of a classical play by Nikolai Gogol about a government inspector arriving into a corrupt town for an incognito investigation.}.
The current implementation supports only Intel x86, which we chose as a worst-case target for our method: a superscalar CPU with several unpatched microarchitectural vulnerabilities, no detailed descriptions of speculation mechanisms, and no direct control over the microarchitectural state.

We evaluated \sys{} on two different microarchitectures, Skylake and Coffee Lake, and with different microcode patches.
We test these targets against a sequence of increasingly permissive contracts.
This gradually filters out common violations, and narrows down on more subtle violations.
The key highlights of our evaluation are:
\begin{enumerate}
    \item When testing a patched Skylake against a restrictive contract that states that speculation exposes \emph{no information}, \sys{} detects a violation within a few minutes.
    Inspection shows the violation stems from the leakage during branch prediction, i.e. a representative of Spectre V1.
    \item When testing Skylake with \emph{V4 patch disabled} against a contract that permits leakage during branch prediction (and is hence \emph{not} violated by V1), \sys{} detects a violation due to address prediction, i.e., a representative of Spectre V4.
    \item When further weakening the contract to permit leaks during both types of speculation, \sys{} still detects a violation.
    This violation is a novel (minor) variant of Spectre where the timing of variable-latency instructions (which is \emph{not} permitted to leak according to the contract) leaks into L1D through a race condition induced by speculation.
    \item When making microcode assists possible during collection of the hardware traces,
    \sys{} surfaces MDS~\cite{RIDL,Fallout} on the same CPU and LVI-Null~\cite{VanBulck2020} on a CPU patched against MDS\@.
    \item When used to validate an assumption that stores do not modify the cache state until they retire, made in recent defence proposals~\cite{STT2019,Wang2020}, \sys{} discovered that this assumption does not hold in Coffee Lake.
    \item In terms of speed, \sys{} processes over 200 test cases per hour for complex contracts, and with several hundreds of inputs per test case, which enables discovery of Spectre V1, V4, MDS, and LVI-Null in under two hours, on average.
\end{enumerate}

\myparagraph{Summary}
In summary, starting from simple contracts, \sys{} could automatically generate gadgets that represent all three of the known types of speculative leakage: speculation of control flow, address prediction, and speculation on hardware exceptions.
Notably (and perhaps surprisingly), \sys{} finds them within only a few hours of testing on an ordinary desktop PC, despite the enormous size of the search space.
The reason is that counterexample search is \emph{not} akin to finding a needle in a haystack.
Instead, microarchitectural leaks manifest in many programs, and it is sufficient to find only one of them.
This result demonstrates the practicality of testing complex real-world CPUs against speculation contracts.

The source code is publicly available under: 

\centerline{\url{https://github.com/hw-sw-contracts/revizor}}


\section{Background: Contracts}
\label{sec:background}

\bk{Add two sentences about what is to come}

\subsection{Hardware Traces and Side-channel Leakage}
\label{subsec:bak-leak}

We consider an abstract side-channel attack model whereby an adversary can use side-channels~\cite{Tromer10,Yarom14,Osvik2006} to extract secret information about a victim program $\program{}$ execution.
Specifically, we focus on microarchitectural side-channels, such as cache timing.
We define a \term{hardware trace} as a sequence of all the observations made through the side-channel after each instruction during a program execution.

We represent the hardware trace as the output of a function $\attack{}$
\begin{equation*}
    \htrace{}=\attack{}(\program{},\archstate{},\microstate{})
\end{equation*}
that takes three input parameters:
(1) the victim program $\program{}$;
(2) the input $\archstate{}$ processed by the victim's program (i.e., the architectural state including registers and main memory);
(3) the microarchitectural context $\microstate{}$ in which it executes.

The information exposed by a hardware trace depends on the assumed side-channel and threat model.
\begin{example}
    If the threat model includes attacks on a data cache, $\htrace{}$ is composed of the cache set indexes used by $\program{}$'s loads and stores.
    If it includes attacks on an instruction cache, $\htrace{}$ contains the addresses of executed instructions.
\end{example}

A program \term{leaks} information via side-channels when its hardware traces depend on the inputs ($\archstate{}$):
We assume the attacker knows $\program{}$ and can manipulate $\microstate{}$, hence any difference between the hardware traces implies difference in  $\archstate{}$, which effectively exposes information to the attacker.

Intuitively, hardware traces encompass the microarchitectural leaks during the program execution on a given CPU, including speculative execution.
For example, the trace will record a sensitive memory access during a branch misprediction, such as the leak exploited in Spectre~\cite{Kocher2018}.

\subsection{Legitimate Exposure as a Contract}
\label{subsec:bak-contract}

We now show how speculation contracts can be used to specify the information legitimately exposed by each instruction.

A \term{speculation contract}~\cite{Guarnieri2021} specifies the information that can be exposed by a CPU during a program execution under a given threat model.
For each instruction in the CPU ISA (or a subset thereof), a contract describes the information exposed by the instruction's (\term{observation clause}) and the externally-observable speculation that the instruction may trigger (\term{execution clause}).
When a contract covers a subset of ISA, the leakage of unspecified instructions is undefined.


\begin{table}[tbp]
    \vspace{0.5cm}
    \center
    \begin{tabular}{l|l|l}
        & Observation    Clause         & Execution     Clause           \\
        \hline
        \hline
        Load        & \code{expose:\ ADDRESS}  & \code{None}                           \\
        \hline
        Store       & \code{expose:\ ADDRESS} & \code{None}                           \\
        \hline
        Cond.  & \code{None} & \code{speculate:} \\
        Jump  &             &  \code{  if(INVERTED\_CONDITION)\{}\\
        & & \code{\ \ IP = IP + TARGET\}} \\
        \hline
        Other        & \code{None}                          & \code{None}                           \\
        \hline
    \end{tabular}
    \vspace{0.2cm}
    \caption{Summary of \memcond{}. Note that the execution clause
    describes the speculative behavior of a conditional jump, as the jump
    takes place (IP is updated) if the condition is false, the opposite of the
    non-speculative execution.}
    \label{tab:memseq}
\end{table}

\begin{example}
    Consider a contract called \memcond{} (summarized in \tabref{memseq}).
    Through the observation clauses of loads and stores, the contract prescribes that addresses of all memory access may be exposed (hence \textit{MEM}).
    The execution clause of conditional branches describes their misprediction, thus the contract prescribes that branch targets may be mispredicted (hence \textit{COND}).
    This way, the contract models a data cache side channel on a CPU with branch prediction.
\end{example}

A \term{contract trace} $\ctrace$ contains the sequence of \emph{all} the observations the contract allows to be exposed after each instruction during a program execution, including the instructions executed speculatively.
Conversely, the information that is \emph{not} exposed via $\ctrace{}$ is supposed to be kept secret.

We represent a contract as a function $\contract{}$ that maps the program $\program$ and its input $\archstate$ to a contract trace $\ctrace$:
\begin{equation*}
    \ctrace{}=\contract{}(\program{},\archstate{})
\end{equation*}

\begin{example}
    Consider the program in~\figref{example-violation}, executed with an input \code{data=\{x=10,y=20\}}.
    The \memcond{} contract trace is \code{ctrace=} \code{=[0x110,0x220]}, representing that the load at line 1 exposes the accessed address during normal execution, and the load at line 3 exposes its address during speculative execution triggered by the branch at line 2.
\end{example}

\begin{figure}[tbp]
  \vspace{1cm}
    \begin{lstlisting}[]
z = array1[x] # base of array1 is 0x100
if (y < 10)
  z = array2[y] # base of array2 is 0x200\end{lstlisting}
    \caption{Example of Spectre V1}
    \label{fig:example-violation}
    \vspace{1cm}
\end{figure}

A CPU complies~\cite{Guarnieri2021} with a contract when its hardware traces (collected on the actual CPU) leak at most as much information as the contract traces.
Formally, we require that whenever any two executions of \emph{any} program have the same contract trace (implying the difference between inputs is not exposed), the respective hardware traces should also match.
\begin{definition}
    \label{def:compliance}
    A CPU \term{complies} with a $\contract$ if, for all programs~$\program$, all input pairs~$(\archstate,\archstate')$, and all initial microarchitectural states~$\microstate$:
    \begin{gather*}
        \contract(\program,\archstate) \!=\! \contract(\program,\archstate')  \\
        \implies \attack(\program,\archstate,\microstate) \!=\! \attack(\program,\archstate',\microstate)
    \end{gather*}\label{eq:comply}
    \vspace{-0.2cm}
\end{definition}

This approach is called \emph{relational} reasoning, and is natural for expressing information flow properties~\cite{ClarksonS10}.
In the corresponding terminology~\cite{sabelfeld2003language}, \defref{compliance} requires that any program that is non-interferent with respect to a contract must also be non-interferent on the CPU\@.

Conversely, a CPU \term{violates} a contract if there exists a program $\program$, a microarchitectural state \microstate, and two inputs $\archstate, \archstate'$ that agree on their contract traces but \emph{dis}agree on the hardware traces.
We call the tuple $(\program,\microstate,\archstate,\archstate')$ a contract \term{counterexample}.
The counterexample witnesses that an adversary can learn more information from hardware traces than what the contract specifies.
A counterexample indicates a potential microarchitectural vulnerability that was not accounted for by the contract.

\begin{example}
    Consider a contract, called \memseq{}, which allows exposure of memory accesses (similarly to \memcond{}), but limits it to only non-speculative accesses.
    A CPU that leaks on speculatively executed branches will violate \memseq{}.
    Its counterexample is the program in~\figref{example-violation} together with inputs \code{data1=\{x=10,y=20\}} and \code{data2=\{x=10,y=30\}} and a context that triggers a misprediction:
    The contract trace for both inputs is \code{ctrace=[0x110]}.
    However, when the CPU mispredicts the branch (line~2) and speculatively accesses memory (line~3), the hardware traces will diverge (\code{htrace1=[0x110,0x220]} and \code{htrace2=[0x110,0x230]}) .
    Yet, this is not a counterexample to \memcond{}, because its contract traces already expose the memory accesses on both paths of a branch.
\end{example}


\subsection{Concrete Contracts of Speculation}
\label{sec:our-contracts}

A contract is constructed from a combination of an observation and execution clauses.
We first describe individual clauses, and then show how they form concrete contracts.

\myparagraphnodot{Observation clauses:}
\begin{itemize}
    \item \textit{MEM (Memory Address)}: exposes the addresses of data loads and stores.
    Represents a data cache timing side-channel attack.
    \item \textit{CT (Constant-Time)}: extends MEM by additionally exposing Program Counter.
    Represents both data and instruction cache attacks.
    Based on a typical threat model for constant-time programming (hence the name), except it does not expose the execution time of variable-latency operations.
    \item \textit{ARCH (Architectural Observer)}: extends CT by additionally exposing the \emph{values} loaded from memory.
    Represents a same-address-space attack, such as assumed in the Speculative
    Taint Tracking paper~\cite{STT2019}.
\end{itemize}

\myparagraphnodot{Execution clauses:}
\begin{itemize}
    \item \seq{}: observations are only collected during \emph{sequential execution} (in-order, nonspeculative).
    This is a model of a processor that allows speculation but constrains the information leaked during the speculation when combined with the appropriate observation clause.
    \item \cond{}: observations are also collected after \emph{conditional jump misprediction}.
    That is, they are collected from both correct and mispredicted paths.
    The length of the mispredicted path is limited by a predefined speculation window.
    \item \bpas{}: observations are collected after \emph{store bypass}: all stores are speculatively skipped.
    The mis-speculated execution rolls back after the speculation window as in \cond{}.
    \item \cond{}-\bpas{}: Combination of \cond{} and \bpas{}.
\end{itemize}

\myparagraph{Full contracts}
We illustrate how the clauses form a contract with examples:

\begin{example}
    \ctcb{} exposes addresses of all memory accesses and of all control-flow transitions,
    including those on mispredicted paths of conditional branches.
    \ctcb{} models a CPU vulnerable to Spectre V1 attacks.
\end{example}

\begin{example}
    \archseq{} exposes addresses and values of non-speculative loads and stores.
    There is a subtle difference from \memseq{}.
    While \memseq{} disallows speculative leakage of any values, \archseq{} disallows leakage of only speculatively loaded values.
    This is equivalent to transient noninterference\cite{STT2019}.
\end{example}


\section{Challenges of Testing Contract Compliance}
\label{sec:challenges}

In this work, we leverage contracts to check compliance of complex commercial CPUs under realistic threat models.
Assuming that a contract properly exposes the expected information leakage in a CPU, finding a counterexample would signify an unexpected, hence potentially exploitable, leakage.

While the original paper~\cite{Guarnieri2021} proved compliance on an \emph{abstract} CPU with toy assembly, testing compliance of a \emph{real hardware} CPU with complex ISA poses significant challenges.

\subsection{How to Find a Counterexample?}
\label{subsec:challenges-complexity}

The search space for counterexamples is all possible programs, inputs, and all microarchitectural contexts.
Such an immense search space cannot be explored exhaustively, thus requiring a targeted search.

\challenge{Binary Generation}
\label{ch:search}
While a contract prescribes which instructions are permitted to speculate and expose information, we search for \emph{unexpected} speculation and leakage, thus we need to collect traces that encompass all the instructions.
Furthermore, a particular \emph{sequence of instructions} is usually required to produce an observable leakage, thus we need to test different instruction sequences.
Moreover, to trigger an \emph{incorrect} speculation (e.g., a branch misprediction), we need to prime the microarchitectural state in diverse ways.
All of it calls for a search strategy that tests diverse instruction sequences with diverse inputs, but with a priority to those that are likely to leak or to produce speculation.

\challenge{Input Generation}
\label{ch:meth-eq-coverage}
For an input to be useful in forming a counterexample, we need another input that produces the same contract trace.
Such inputs are called \emph{effective inputs}.
The \emph{in}effective inputs which produce a unique contract trace constitute a wasted effort as they cannot, by definition, reveal contract violation.
This challenge calls for a more structured input generation approach rather than a simple random one, as the probability that multiple random inputs will produce the same contract trace is low.

\subsection{How to Get Stable Hardware Traces on a Real CPU?}
\label{subsec:challenges-htraces}

\challenge{Collection of Hardware Traces}
\label{ch:meth-htraces}
CPUs have no direct interface to record information leaked in hardware traces, such as addresses accessed in a speculative path.
Thus, we have to perform indirect sampling-based measurements, which are inevitably imprecise and incomplete.

\challenge{Uncontrolled Microarchitectural State}
\label{ch:meth-nondeterminism}
Black-box CPUs normally have no direct way to set the microarchitectural context for test execution as required by~\defref{compliance}.
For example, branch predictors are not accessible architecturally, and some are not even disclosed.
Moreover, speculation depends on multiple, often unknown factors, such as fine-grained power saving~\cite{naveh2006power,rotem2019mechanism}, or contention on shared resources.
Thus, speculation can happen nondeterministically, and cause divergent traces without a real information leak (false positive).
On the other hand, if the speculation is never triggered during the measurement, speculative leaks cannot be observed, leading to false compliance (false negative).

\challenge{Noisy Measurements}
\label{ch:meth-noise}
The measurements are influenced by neighbour processes on the system, by hardware mechanisms (e.g., prefetching), and by inherent imprecision of the measurement tools (e.g., timing measurements).
This challenge differs from~\chref{meth-nondeterminism} as it affects the measurement precision rather than the program execution.
The noise may result in divergence between the otherwise equivalent traces, leading to a false positive.

\subsection{How to Produce Contract Traces?}
\label{subsec:challenges-ctraces}

\challenge{Collection of Contract Traces}
\label{ch:ctraces}
All contracts in~\cite{Guarnieri2021} are defined for a toy assembly;
it is unclear how to collect traces for a contract describing a complex ISA\@.
To allow realistic compliance check, we need work with real binaries generated via standard compiler tool chain.
Hence, we need a method to automatically collect contract-prescribed observations for a given program executed with a given input.


\section{\methodlong{}}
\label{sec:method}

\begin{figure}[t!]
    \centering
    \includegraphics[width=\columnwidth]{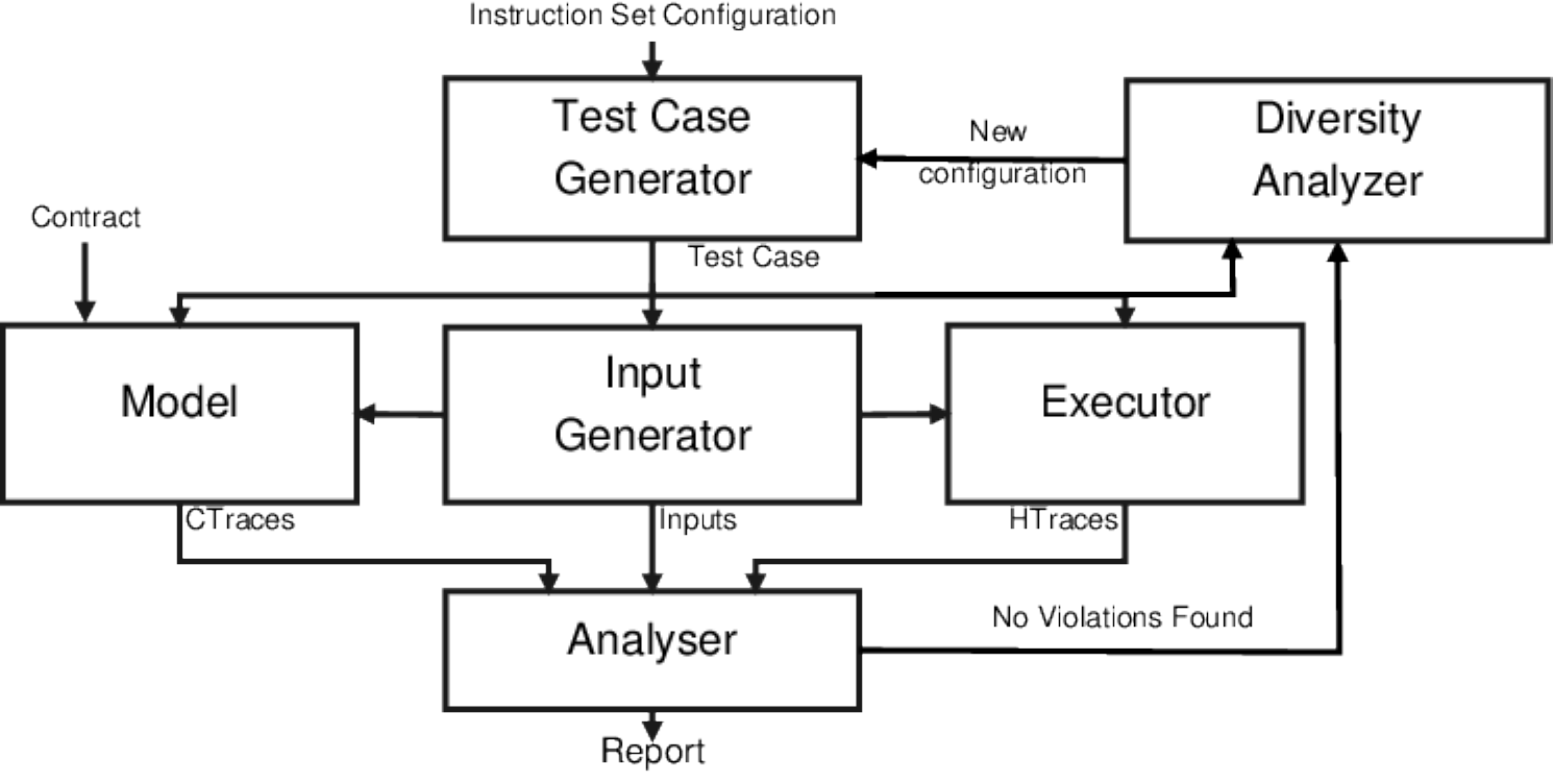}
    \caption{Main flow of \methodlong{}.}
    \label{fig:arch}
\end{figure}

We present \methodlong{} (\method{}), our approach to identifying contract violations in black-box CPUs.
Here we provide a high-level description, with the technical details to follow (\secref{design}).
\figref{arch} shows the main steps.

\myparagraph{Test case and input generation}
We sample the search space of programs, inputs and microarchitectural states to find counterexamples.
The generated instruction sequences (\term{test cases}) are comprised of the ISA subset described by the contract.
The test cases and respective inputs to them are generated to achieve high diversity and to increase speculation or leakage potential (\subsecref{des-testgen} and \subsecref{des-input}).

\myparagraph{Collecting contract traces}
We implement an executable Model of the contract to allow automatic collection of contract traces for standard binaries.
For this, we modify a functional CPU emulator to implement speculative control flow based on a contract's execution clause, and to record traces based on its observation clause (\subsecref{des-model}).

\myparagraph{Collecting hardware traces}
We collect hardware traces by executing the test case on the CPU under test and measuring the observable microarchitectural state changes during the execution according to the threat model.
The executor employs several methods to achieve consistent and repeatable measurements (\subsecref{des-executor}).

\myparagraph{Relational Analysis}
We analyze the contract and hardware traces to identify violations of~\defref{compliance}.
This requires \emph{relational} reasoning:

\begin{enumerate}
    \item We partition inputs into groups, which we call \term{input classes}.
    All inputs within a class have the same contract trace.
    Thus, input classes correspond to the equivalence classes of equality on contract traces.
    Classes with a single (ineffective) input are discarded.
    \item For each class, we check if all inputs within a class have the same hardware trace.
\end{enumerate}

If the check fails on any of the classes, we found a counterexample that
witnesses contract violation (\subsecref{des-analyser}).

\myparagraph{Diversity-guided generation}
The testing process is performed in rounds, where earlier rounds exercise smaller search space (i.e., shorter instruction sequences, fewer basic blocks) to speed up testing.
After each round that did not yield a counterexample, we invoke a test case diversity analysis which may trigger reconfiguration of the test generator to produce richer test cases, gradually expanding the search space (\subsecref{coverage}).


\section{Design and Implementation}
\label{sec:design}

We build a tool \sys{} that implements \method{} for practical end-to-end testing of x86 CPUs against speculation contracts.
We describe the individual components of \sys{} and how they address the challenges outlined in~\secref{challenges}.

\subsection{Test Case Generator}
\label{subsec:des-testgen}

The task of the test case generator is to sample the search space of all possible programs.
As described in~\chref{search}, the sampling should be diverse, so that we have a chance to observe an \emph{unexpected} leakage or speculation.
Fully random generation, however, might lead to generating incorrect programs, e.g., with invalid control flow or memory accesses, leading to unhandled exceptions during their execution.
This is why we rely on a randomized generation algorithm which imposes a certain structure on the generated instruction sequence and its memory accesses.
It works as follows:
\begin{enumerate}
    \item Generate a random Directed Acyclic Graph (DAG) of basic blocks;
    \item Add jump instructions (terminators) at the end of basic block to ensure the control flow matches the DAG.
    \item Add random instructions from the tested ISA subset;
    \item Instrument instructions to avoid faults:
    \begin{enumerate}
        \item mask memory addresses to confine them within a dedicated memory region, which we call \emph{sandbox};
        \item modify division operands to avoid division by zero;
    \end{enumerate}
    \item Compile the test case into a binary.
\end{enumerate}

The total number of instructions, functions, and basic blocks per test, as well as the tested instruction (sub)set are specified by the user.
We borrow the ISA description from nanoBench~\cite{Abel2019a}.


\begin{figure}[t]
    \footnotesize
    \begin{lstlisting}[frame=t]
OR RAX, 468722461
\end{lstlisting}\begin{lstlisting}[frame=none,backgroundcolor=\color{gray!10},firstnumber=2]
AND RAX, 0b111111000000
\end{lstlisting}\begin{lstlisting}[frame=none,firstnumber=3]
LOCK SUB byte ptr [R14 + RAX], 35
JNS .bb1
JMP .bb2
\end{lstlisting}\begin{lstlisting}[frame=none,backgroundcolor=\color{gray!10},firstnumber=6]
.bb1: AND RCX, 0b111111000000
\end{lstlisting}\begin{lstlisting}[frame=none,firstnumber=7]
REX SUB byte ptr [R14 + RCX], AL
CMOVNBE EBX, EBX
OR DX, 30415
JMP .bb2
.bb2: AND RBX, 1276527841
\end{lstlisting}\begin{lstlisting}[frame=none,backgroundcolor=\color{gray!10},firstnumber=12]
AND RDX, 0b111111000000
\end{lstlisting}\begin{lstlisting}[frame=b,firstnumber=13]
CMOVBE RCX, qword ptr [R14 + RDX]
CMP BX, AX\end{lstlisting}
    \caption{Randomly generated test case}
    \label{fig:test-case-raw}
\end{figure}

\begin{example}
    \figref{test-case-raw} shows a test case example, produced in multiple steps:
    \circled{1} The generator created a DAG with three nodes. 
    \circled{2} Connected the nodes by placing either conditional or direct jumps (lines 4--5, 10).
    \circled{3} Added random instructions until a specified size was reached (lines 1, 3, 7--9, 13, 14).
    \circled{4} Masked the memory accesses and aligned to the sandbox base in \text{R14} (lines~2, 6, 12).
\end{example}

We use DAG as a basis for the generation process to confine the control flow and avoid infinite loops.
The limitation of this approach is that we do not test loops, which may prevent \sys{} from detecting loop-based contract violations.
However, it is only a technical limitation and, in the future, it could be solved by analyzing the control flow of test cases and enforcing loop termination at generation time.

\myparagraph{Improving input effectiveness}
Using many hardware registers and larger sandbox results in low input effectiveness (\chref{meth-eq-coverage}), as it increases the likelihood of unique contract traces that cannot be used for relational testing.
To improve input effectiveness, the generator generates programs with only four registers, confines the memory sandbox to one or two 4K memory pages, and aligns memory accesses to a cache line (64B).
To test different alignments, the accesses are further offset by a random value between 0 and 64 (the same within a test case but different across test cases).

\subsection{Input Generator}
\label{subsec:des-input}

An input is a set of values to initialize the architectural state, which includes registers (including \texttt{FLAGS}) and the memory sandbox.
\sys{} creates random inputs with a 32-bit PRNG\@.

The initial number of inputs per test case is configured up-front, and it increases every time the diversity analyser triggers a reconfiguration (\subsecref{coverage})

\myparagraph{Improving input effectiveness}
Higher entropy of the PRNG leads to lower input effectiveness (\chref{meth-eq-coverage}), because the probability of finding colliding contract traces decreases.
We amend this issue by artificially reducing the PRNG entropy by masking some output bits;
lower entropy results in higher input effectiveness but smaller range of tested values.
We expect that more sophisticated techniques for creating inputs (e.g., based on symbolic execution) would be able to achieve high effectiveness without manipulating the PRNG\@.

\subsection{Executor}
\label{subsec:des-executor}

The executor has three tasks: (1) collect hardware traces when executing test
cases on the CPU (\chref{meth-htraces}), (2) set the microarchitectural context for the execution (\chref{meth-nondeterminism}), and (3) eliminate measurement noise (\chref{meth-noise}).

\myparagraph{Collecting hardware traces}
To collect traces we employ methods used by side-channel attacks, but in a fully controlled environment.
This allows us to record hardware traces corresponding to the measurements of a powerful worst-case attacker, and spot all consistently-observed leaks via the microarchitectural state.
The process involves the following steps:
\begin{enumerate}
    \item Load the test case into a dedicated region of memory,
    \item Set memory and registers according to the inputs,
    \item Prepare the side-channel (e.g., prime cache lines),
    \item Invoke the test case,
    \item Measure the microarchitectural changes (e.g., probe cache lines) via the side-channel, thus producing a trace.
\end{enumerate}

The measurement (steps 2--5) repeats for all inputs, thus producing a hardware trace for each test case-input pair.

Our implementation supports several measurement modes:
\begin{itemize}
    \item \emph{Prime +Probe}~\cite{Osvik2006}, \emph{Flush +Reload}~\cite{Yarom14}, and \emph{Evict +Reload}~\cite{Gruss2015} modes use the corresponding attack on L1D cache.
    \item In \emph{*+Assist} mode, the executor includes microcode assists.
    It clears the ``Accessed'' bit in one of the accessible pages such that the first store or load triggers an assist~\footnote{Microcode assist is a situation when the CPU redirects the control to an internal microcode routine to execute a complex operation, such as setting a page table bit.}.
\end{itemize}

\begin{example}
    The hardware trace corresponding to running executor in L1D Prime+Probe mode is a sequence of bits, each representing whether a specific cache set was accessed by the test case or not.
    E.g., the following trace indicates observed memory accesses to sets 0,4,5:
    \texttt{10001100000000000000000000000000}
\end{example}

\myparagraph{Setting the microarchitectural context}
We cannot directly control the microarchitectural context before the test execution (\chref{meth-nondeterminism}).
To deal with this, we develop a technique called \term{priming}, where we collect traces for a large number of pseudorandom inputs (\subsecref{des-input}) to the same test case \emph{in a sequence}.
In this way, execution with one input effectively sets the microarchitectural context for the next input.
This enables collection of hardware traces with predictors primed in diverse but deterministic fashion, which is key to obtaining traces that are stable enough for equality checks.

Yet priming may result in undesirable artifacts.
For this, recall that \method{} searches for inputs $\archstate{}_1$ and $\archstate{}_2$ from the same input class, but with divergent hardware traces:
\begin{equation*}
    \attack{} ( \program{},\archstate{}_1,\microstate{}) \neq \attack{}(\program{},\archstate{}_2,\microstate{})
\end{equation*}
Due to priming, however, the contexts for each input is different, and the actual equality check is:
\begin{equation*}
    \attack{} ( \program{},\archstate{}_1,\microstate{}_1) \neq \attack{}(\program{},\archstate{}_2,\microstate{}_2)
\end{equation*}
Therefore, the divergence of traces could be caused by differences in the microarchitectural contexts $\microstate{}_1$ and $\microstate{}_2$.
For example, earlier inputs can train branch predictors in a way that would prevent speculation for the latter inputs.

To filter such cases and verify that the divergence is caused by inputs and by contexts, we swap $\archstate{}_1$ and $\archstate{}_2$ in the priming sequence, which enables us to test $\archstate{}_1$ with the context $\microstate{}_2$ and vice versa.
That is, we test the following:
\begin{gather*}
    \attack{}(\program{},\archstate{}_1,\microstate{}_2) = \attack{}(\program{},\archstate{}_2,\microstate{}_2) \\
    \land \;
    \attack{}(\program{},\archstate{}_1,\microstate{}_1) = \attack{}(\program{},\archstate{}_2,\microstate{}_1)
\end{gather*}  %
If this condition holds, we discard the divergence as a measurement artifact, otherwise we report a contract violation.

\begin{example}
    Consider two inputs with the same contract trace but different hardware traces; in the original sequence of inputs, the first was at position 100 ($i_{100}$) and the second at 200 ($i_{200}$).
    For priming, the executor tests sequences $(i_1\dots i_{99}, i_{200}, i_{101} \dots i_{199}, i_{200})$ and $(i_1 \dots i_{99}, i_{100}, i_{101} \dots i_{199}, i_{100})$.
    The executor will consider it a false positive if $i_{100}$ at position 200 produces the same trace as $i_{200}$ at position 200, and vice versa.
\end{example}

\myparagraph{Eliminating measurement noise}
Hardware traces in the same input class may also diverge (and thus incorrectly considered as contract violation) due to several additional sources of inconsistencies which we eliminate as follows:

\begin{enumerate}
    \item \emph{Eliminating measurement noise} (\chref{meth-noise}).
    The executor uses performance counters for cache attacks by reading the L1D miss counter before and after probing a cache line.
    It proved to give more stable results than timing readings.
    \item \emph{Eliminating external software noise} (\chref{meth-noise}).
    We run the executor as a kernel module (based on nanoBench~\cite{Abel2019a}).
    A test is executed on a single core, with hyperthreading, prefetching, and interrupts disabled.
    The executor also monitors System Management Interrupts (SMI)~\cite{intelsys} to discard those measurements polluted by an SMI\@.
    \item \emph{Reducing nondeterminism} (\chref{meth-nondeterminism}).
    We repeat each measurement (50 times in our experiments) after several rounds of warm-up, and discard one-off traces as likely caused by noise.
    We then take the union of all traces collected from the executions of a test case with \emph{the same input}, which encompasses all consistently observed variants of speculative behavior under different microarchitectural contexts.
\end{enumerate}

\begin{example}
    Consider again the test case in~\figref{test-case-raw}.
    If the branch in line~6 is speculated differently across the runs, one input may produce different traces:

    \texttt{00001010000001000000000000000001}

    \texttt{00001000000001000000000000000001}

    The first trace is with a misprediction (cache set 7), and the second without.
    The merged trace is:

    \texttt{00001010000001000000000000000001}
\end{example}

Discarding all outliers observed only once during the test might miss rare cases that reveal real leaks.
However, we found it necessary from the practical perspective: each reported violation requires manual investigation.
Since the outliers turned out to be notoriously hard to reproduce and verify, we opted to focus on the leaks that are easier to distinguish from the noise.

\subsection{Model}
\label{subsec:des-model}

Model's task is to automate the collection of contract traces (\chref{ctraces}).
We achieve this by executing test cases on an ISA-level emulator modified according to the contract.
The emulator implements the contract's execution clause, such as exploring all
speculative execution paths, followed by a rollback, and it collects observations based on the observation clause.
The resulting trace is a list of observations collected while executing a test case with a single input.
We base our implementation on Unicorn~\cite{quynh2015unicorn}, a customizable x86 emulator, modified to implement the clauses listed in~\secref{our-contracts}.

\myparagraph{Observation Clauses}
When the emulator executes an instruction listed in the observation clause, it records its exposed information into the trace.
This happens during both normal and speculative execution, unless the contract states otherwise.

\begin{example}
    Consider the test case \figref{test-case-raw} and the contract \memseq{}.
    As prescribed by the contract, the model records the accessed
    addresses when executing lines~3, 7, 13 (\figref{test-case-raw}).
    Suppose, the branch (line~4) was not taken; the store (line~3) accessed \code{0x100}; and the load (line~13) accessed \code{0x340}.
    Then, the contract trace is \code{ctrace=[0x100,} \code{0x340]}.
\end{example}

\myparagraphnodot{Execution Clauses}
are implemented similarly to the speculation exposure mechanism introduced in SpecFuzz~\cite{Oleksenko2020}:
Upon encountering an instruction with a non-empty execution clause (e.g., a branch in \memcond{}), the emulator takes a checkpoint.
The emulator then simulates speculation as described by the clause until (1) the test case ends, (2) the first serializing instruction is encountered, or (3) the maximum possible speculation depth is reached\footnote{The speculation depth is a configurable parameter. In our experiments, we used 250 instructions, based on the ROB size in Skylake CPUs.}.
Then, it rolls back and continues normal execution.

As multiple mispredictions may happen together, the emulator supports nested speculation through a stack of checkpoints:
When starting a simulation, the checkpoint is pushed, and afterwards, the emulator rolls back to the topmost checkpoint.

Practically, however, nested speculations greatly reduce the testing speed, which is why we disable nesting by default.
This artificially reduces the amount of permitted leakage by the contract, potentially causing false violations (since hardware traces would still include nested speculations).
To identify such false violations, \sys{} re-executes all reported violations with nesting enabled.

\subsection{Analyzer}
\label{subsec:des-analyser}

The analyzer compares traces by using relational analysis (\secref{method}).
As hardware traces are obtained as the union of observations collected from the same input in different microarchitectural contexts (\subsecref{des-executor}), we relax requiring equality of hardware traces to requiring only a subset relation.
Specifically, we consider two traces equivalent if every observation included in one trace is also included in the other trace.

The intuition behind the heuristic is as follows.
If the mismatch is caused by an inconsistent execution of a speculative path among the inputs, one of the inputs executed fewer instructions, therefore fewer observations would appear in the trace, but those that appear match.
In contrast, if the mismatch is caused by a secret-dependent instruction, the traces contain the same number of observations, but their values differ.
To validate this intuition, we manually examined multiple such examples and did not observe any real violation.

\subsection{Test Diversity Analysis}
\label{subsec:coverage}

If a testing round did not detect any violation, we need to decide how to improve the chances of finding one in the next round.
As we test black-box CPUs we cannot measure coverage of the exercised CPU features to guide the test generation in the next round.

Instead, we seek to estimate the likelihood to exercise new speculative paths with the current configuration of the test case generator by analyzing the diversity of the tests we ran so far (\chref{search}).
We capture diversity tests using a new measure called \emph{pattern coverage}, which counts data and control dependencies that are likely to cause pipeline hazards.
We expect higher pattern coverage to correlate with higher chances to surface speculative leaks.
Therefore, if a testing round does not improve the pattern coverage of the tests so far, new speculative paths are unlikely to be explored.
To facilitate generation of more diverse tests, \sys{} then increases the number of instructions and basic blocks per test.
We now discuss this approach in more details.

\myparagraph{Patterns of instructions}
We define patterns in terms of instruction \emph{pairs}.
To simplify the counting of pattern coverage we require that the instructions are consecutive, which corresponds to the worst case for creating hazards.
We distinguish three types:

\begin{enumerate}
    \item  A memory dependency pattern is two memory accesses to the same address.
    We consider 4 patterns: store-after-store, store-after-load, load-after-store, load-after-load.
    \item A register dependency pattern is when one instruction uses a result of another instruction.
    We consider 2 patterns: dependency over a general-purpose register, and over \code{FLAGS}.
    \item A control dependency pattern is an instruction that modifies the control flow followed by any other instruction.
    In this paper we consider 2 patterns: conditional and unconditional jumps.
    Larger instruction sets may include indirect jumps, calls, returns, etc.
\end{enumerate}

We say that a program with an input \emph{matches} a pattern if that pattern is found in two consecutive instructions in the corresponding instruction stream.
Since a single input cannot form a counterexample, a pattern is \emph{covered} if a program and two inputs in the same input class match the pattern.


\begin{table*}[tbp]
    \center
    \begin{tabular}{l||c|c|c|c|c|c||c|c}
        & Target 1 & Target 2 & Target 3 & Target 4 & Target 5 & Target 6 & Target 7 & Target 8 \\
        \hline
        \hline
        CPU & \multicolumn{6}{c||}{Skylake} & Skylake & \multicolumn{1}{c}{Coffee Lake} \\
        \hline
        V4 patch & \multicolumn{3}{c|}{off} & \multicolumn{3}{c||}{on} & \multicolumn{2}{c}{on} \\
        \hline
        Instruction Set & \ibasic & \ibm & \ibmv & \ibmv & \ibcm & \ibcmv & \multicolumn{2}{c}{\ibm} \\
        \hline
        Executor Mode & \multicolumn{6}{c||}{Prime+Probe} & \multicolumn{2}{c}{Prime+Probe+Assist} \\
        \hline
    \end{tabular}
    \caption{Description of the experimental setups.}
    \label{tab:targets}
    \vspace{-0.2cm}
\end{table*}

\begin{table*}[tbp]
    \center
    \begin{tabular}{l|l|l|l|l|l|l|l|l}
        & Target 1       & Target 2       & Target 3                 & Target 4       & Target 5       & Target 6                 & Target 7       & Target 8            \\
        \hline
        \hline
        \ctseq   & $\times${}   & \checkmark{} (V4)  & \checkmark{} (V4)            & $\times${}   & \checkmark{} (V1)  & \checkmark{} (V1)           & \checkmark{} (MDS)         & \checkmark{} (LVI-Null)    \\
        \ctsbp   & $\times$$^*$ & $\times$     & \checkmark{} (V4-var$^{**}$) & $\times$$^*$ & \checkmark{} (V1) & \checkmark{} (V1)        & \checkmark{} (MDS)         & \checkmark{} (LVI-Null)    \\
        \ctcb    & $\times$$^*$ & \checkmark{} (V4)  & \checkmark{} (V4)            & $\times$$^*$ & $\times$     & \checkmark{} (V1-var$^{**}$)     & \checkmark{} (MDS)         & \checkmark{} (LVI-Null)    \\
        \ctcbsbp & $\times$$^*$ & $\times$$^*$ & \checkmark{} (V4-var$^{**}$) & $\times$$^*$ & $\times$$^*$    & \checkmark{} (V1-var$^{**}$)     & \checkmark{} (MDS)              & \checkmark{} (LVI-Null) \\
        \hline
        \multicolumn{6}{l}{\footnotesize{$^*$ we did not repeat the experiment as a stronger contract was already satisfied.}}  & \multicolumn{3}{l}{\footnotesize{$^{**}$ the violation represents a novel speculative vulnerability.}} \\
    \end{tabular}
    \caption{Testing results.
        \checkmark{} means \sys{} detected a violation; $\times${} means \sys{} detected no violations within 24h of testing.
        In parenthesis are Spectre-type vulnerabilities revealed by the detected violations.}
    \label{tab:main-fuzz}
\end{table*}

To provide opportunities for interaction between different speculation types, we count not just individual patterns, but also their combinations.

\myparagraph{Implementation}
We implement tracking of patterns as part of the Model (\subsecref{des-model}):
While collecting contract traces of a test case, the model also records the executed instructions and the addresses of memory accesses.
These data are later analyzed to find the patterns in the instruction streams.

\myparagraph{Coverage Feedback}
We use pattern combination coverage as feedback to the test generator.
We begin with test cases of size $n$ and with at most $m$ basic blocks, tested with $k$ inputs (e.g., 10 instructions, 2 blocks, 50 inputs per test case).
We continue until all individual patterns are covered.
Then, we increase the sizes by constant factors (e.g., 15 instructions, 3 blocks, 75 inputs), and continue testing until all combinations of $2$ patterns are covered, and so on.

\subsection{Postprocessor}
\label{subsec:des-post}

When a violation is detected, the test case is passed to the postprocessor, which
minimizes the test case in three stages:

First, the postprocessor creates a minimal input sequence:
It removes inputs until it finds the smallest sequence to correctly prime the microarchitectural state for the violation.
Second, it creates a minimal test case:
It removes one instruction at a time while checking for violations.
Third, it minimizes the speculative part:
It adds \texttt{LFENCE}s, starting from the last instruction, while checking for violations.
The resulting region without fences is the location of leakage.


\begin{figure}[t]
    \begin{lstlisting}[frame=t]
AND RAX, 0b111111000000 ; LFENCE
\end{lstlisting}\begin{lstlisting}[frame=none,backgroundcolor=\color{gray!10},firstnumber=2]
LOCK SUB byte ptr [R14 + RAX], 35
JNS .bb1 ; LFENCE
\end{lstlisting}\begin{lstlisting}[frame=none,firstnumber=4]
JMP .bb2
\end{lstlisting}\begin{lstlisting}[frame=none,backgroundcolor=\color{gray!10},firstnumber=5]
.bb1: AND RCX, 0b111111000000
REX SUB byte ptr [R14 + RCX], AL
\end{lstlisting}\begin{lstlisting}[frame=b,firstnumber=7]
.bb2: LFENCE\end{lstlisting}
    \caption{Minimized test case, representative of Spectre V1.}
    \label{fig:test-case-min}
\end{figure}

\begin{example}
    \figref{test-case-min} is a minimized version of \figref{test-case-raw}.
    The highlighted region without \texttt{LFENCE}s is the location of leakage:
    The store (line~2) delays the jump (line~3), thus sufficiently prolonging the speculation.
    The jump mispredicts and goes to line~5.
    This causes a speculative execution of \texttt{SUB} (line~6), which has a memory operand and thus leaks the value of \texttt{RCX}.
\end{example}


\section{Evaluation}
\label{sec:evaluation}

In this section, we demonstrate \sys{}'s ability to expose contract violations and automatically identify speculative execution vulnerabilities in two generations of Intel CPUs.

\subsection{Experimental Setup}
\label{subsec:eval-setups}

We test multiple CPUs, ISA subsets, and threat models against several contracts.
The experiments are summarized in \tabref{targets}.

\myparagraph{CPUs (rows 1 and 2)}
We run our experiments on two machines.
The first has Intel Core i7-6700 CPU (Skylake), the second an Intel Core i7-9700 CPU (Coffee Lake).
We analyze Skylake with Spectre V4 microcode patch enabled and disabled.
Coffee Lake has a hardware MDS patch.

\myparagraph{Instruction Sets (row 3)}
We build our test cases from the following subsets of x86\footnote{We do not consider bit count, bit test, and shift instructions because Unicorn sometimes emulates them incorrectly.}:
\begin{itemize}
    \item \ibasic: in-register arithmetic, including logic and bitwise;
    \item \texttt{MEM}: memory operands and loads/stores;
    \item \texttt{VAR}: variable-latency operations (divisions).
    \item \texttt{CB}: conditional branches;
\end{itemize}
This totals in the following number of unique instructions:
\ibasic---325; \ibm---678; \ibmv---687; \ibc---359; \ibcm---710, \ibcmv---719.

We select these particular subsets of instructions to structure the description of results.
As we will see next, each of them surfaces a different type of contract violations.

\myparagraph{Threat Models (row 4)}
We tested contracts against two threat models, \emph{Prime+Probe} and \emph{Prime+Probe+Assist} (see~\subsecref{des-executor}).
Note that Flush/Evict+Reload would produce equivalent traces, as we use a 4KB sandbox, and the 64 L1D cache sets (observed by P+P) correspond to 64 memory blocks in a 4KB region (observed by F+R).

\myparagraph{Configuration}
Generation started from 8 instructions, 2 memory accesses, and 2 basic blocks per test case; 2 bits of input entropy; 50 inputs per test case.
The parameters increased over testing rounds.

\subsection{Testing Results}
\label{subsec:eval-fuzz}

We report our findings when testing the targets in~\tabref{targets} against different contracts.
We tested each for 24 hours or until the first violation was found.
The results are in \tabref{main-fuzz}.

\myparagraph{Target 1: Baseline}
As a baseline, we test the most narrow instruction set \ibasic{} containing only arithmetic operations on Skylake (with V4 patch disabled) using the weakest threat model (P+P without assists).
We expect the target to comply with the most restrictive contract (\ctseq).
The experiments confirm it:
\sys{} did not detect violations (column 1 of~\tabref{main-fuzz}).
Since other contracts are more liberal, the target also complies with more liberal contracts.
This experiment shows that \sys{} successfully mitigates measurement noise and filters the artifacts of non-deterministic execution, producing no false violations.

\myparagraph{Target 2: Memory Accesses}
When augmenting the instruction set with memory accesses to \ibm{} (for the same CPU and threat model), \sys{} detects violations of \ctseq{} and \ctcb{}.
Upon manual inspection, we identify those violations as representative of Spectre V4 (Speculative Store Bypass)~\cite{Google2018}.
\sys{} does not detect violations of \ctsbp{} and \ctcbsbp{}, which is expected as they permit the store bypass\footnote{During the Artifact Evaluation process, 
\sys{} discovered an unexpected counterexample in this experiment, where Target 2 violates \ctsbp{}. It is described in \subsecref{artifact-new-variant}}.

\myparagraph{Target 3: Variable-latency Instructions}
When further augmenting the instruction set with divisions (the only variable-latency instructions in the base x86~\cite{abel2019uops}) to \ibmv, \sys{} finds violations of \ctsbp{} and \ctcbsbp{}.
Upon inspection, they reveal a \emph{novel variant of Spectre V4} that leaks the timing of division (\emph{not} permitted to be exposed according to the contract).
We discuss this variant in~\subsecref{eval-new}.

\myparagraph{Target 4: V4 Patch}
We change the experiment described in Target 3 by enabling the V4 patch on Skylake.
Our experiments do not surface any contract violations, showing that the V4 patch is effective, also against the novel V4 variant.

\myparagraph{Targets 5--6: Conditional Branches}
When augmenting \ibm{} with conditional branches to \ibcm{}, \sys{} detects violations of \ctseq{} and \ctsbp{}. 
Upon inspection, these are representative of Spectre V1~\cite{Kocher2018}.
\sys{} detects no violations of \ctcb{} and \ctcbsbp{}, which is expected as the contracts permit exposing accesses during the execution of a mispredicted branch.

When further augmenting the instruction set with variable-latency instructions to \ibcmv, \sys{} detects violations of \ctcb{} and \ctcbsbp{}.
Similar to Target 3, the violations represent novel variants of Spectre V1.

\myparagraph{Target 7: Microcode Assists}
We now perform experiments with a different threat model, corresponding to an adversary that can cause microcode assists.
To test the assists in isolation, we test \ibm{}, and we enable V4 patch to avoid violations caused by V4.
\sys{} now detects violations of all contracts, which we identify as representative of MDS~\cite{ZombieLoad,Fallout}. 

\myparagraph{Target 8: MDS Patch}
We repeat the experiment in Target 7, but now on Coffee Lake, which has a hardware MDS patch.
\sys{} detected violations on it as well, which we identify as LVI-Null~\cite{VanBulck2020}, a known vulnerability of the MDS patch. 

\myparagraph{Summary}
We see that \sys{} successfully discovered several known and also unknown vulnerabilities, fully automatically, without manual intervention.

\subsection{Novel Variants Discovered}
\label{subsec:eval-new}

\sys{} discovers two new types of speculative leakage of the instruction latency.
As they represent variations on Spectre V1 and V4, and the existing defences prevent them, we did not report them to Intel.
Yet they should be considered when developing future defences, hence we describe them next.

The latency of some operations (e.g., division) depends on their operand values.
The timing difference exposes the values to the attacker who can measure the program's execution time.
However, as \sys{} discovered, the timing can also impact the cache state, thus leaking through caches as well.

\figref{v1-var} shows a simplified version of the V1 variant. 
The key observation is that leakage happens due to a race condition:
\begin{itemize}
    \item if the variable-latency operation (line 1) is faster than the branch instruction (line~2), then the memory access (line~3) could leave a speculative cache trace.
    \item otherwise, the speculation will be squashed before the operation completes, and the memory access will not be executed.
\end{itemize}
As such, the hardware traces expose not only the accessed address, but also the latency of the operation at line~1.

The discovered V4 variant exploits the same race condition;
we expect it to affect other speculative vulnerabilities as well.

\begin{figure}[tbp]
        \begin{lstlisting}[]
b = variable_latency(a)
if (...)  # misprediction
  c = array[b] # executed if the latency is short\end{lstlisting}
    \caption{New Spectre V1 variant (V1-Var), found by \sys{}.}
    \label{fig:v1-var}
\end{figure}

\subsection{Validating Assumptions about Speculative Execution}
\label{subsec:validating-assumptions-about-speculative-execution}

Several defence proposals (STT~\cite{STT2019}, KLEESpectre~\cite{Wang2020}) assume that stores do not modify the cache state until they retire.
We use \sys{} to validate this assumption.
We modify \ctcb{} to capture this assumption in the contract trace, and test our CPUs against it. 
\sys{} discovers no violations in Skylake, but finds a counterexample on Coffee Lake. 
It looks similar to Spectre V1, except the trace is left by a speculative \emph{store}.
This is an evidence that the assumption is wrong and speculative stores \emph{can} modify the cache state. 
Notably, this result has been predicted by previous work, CheckMate~\cite{Trippel2018}.

\subsection{Detection Time}
\label{subsec:eval-detect}


\begin{table}[tbp]
    \small
    \center
    \begin{tabular}{l|c|c|c|c}
        Contract-& \multicolumn{4}{c}{Detection time} \\
        permitted & V4-type       & V1-type     & MDS-type    & LVI-type     \\
        leakage     & (Target 2)     & (Target 5)   & (Target 7)   & (Target 8)    \\
        \hline
        \hline
        None      & 73m 25s (.7)  & 4m 51s (.9) & 5m 35s (.7) & 7m 40s (1.1) \\
        V4        & N/A           & 3m 48s (.7) & 6m 37s (.8) & 3m 06s (1.0) \\
        V1        & 140m 42s (.6) & N/A         & 7m 03s (.8) & 3m 22s  (.3)   \\
        \hline
    \end{tabular}
    \caption{Detection time: the testing time elapsed before the first detected violation.
     The numbers are mean over 10 measurements; in parentheses are coefficients of variation.
     Most vulnerabilities are automatically detected within minutes. The second
     and third rows show that the detection is fast even with multiple leakage
     types in a test case (details in \subsecref{eval-detect}).}
    \label{tab:detection}
\end{table}

We next measure the time required to find a counterexample.
We test each of the targets in~\subsecref{eval-fuzz} that had violations (Targets 2, 5, 7, 8)
\footnote{We did not measure the detection time of the variants discovered in Targets 3 and 6 as they are too rare for repeated measurements.} against \ctseq{} for 10 times and report the average time until the first violation (row 1 of \tabref{detection}).
\ms{I would suggest we put a concrete number for those missing, and then explain that it was not averaged. Otherwise it looks really weird.}

\sys{} detected most violations in under 10 minutes while still using short test cases \ms{it would be great to know the actual numbers}.
This demonstrates \al{not necessarily - correlation does not mean causation. I'm not sure if the detection time would be driven if we had fixed-size test cases.} the importance of diversity-driven feedback.
\sys{} took longer to find V4-like violations as they require a longer speculation window, and the hardware predictor is less prone to misprediction.

\myparagraph{Coping with multiple types of speculation leakages}
We measure how fast \sys{} detects a violation when two types of speculative leakage are present in the test case, but one of them is permitted by the contract;
that is, \sys{} has to detect an unexpected leakage while ignoring an expected leakage.
The second and the third row shows the detection time when Spectre V1 and V4 respectively are permitted by the contract and are present in the test case, while testing against \ctsbp{} (V4 patch disabled).
We observe that these additional leakages did not hinder detection of the vulnerabilities, albeit sometimes slowing down the detection because the model has to execute speculative paths, and having more observations reduces input effectiveness.

\myparagraph{Number of Inputs to Violation}
We analyze the number of random inputs that are required to surface a violation of \ctseq{} with manually-written test cases representing Spectre and MDS vulnerabilities.
\tabref{known-vulns} reports an average of 100 experiments, each with a different input generation seed.
\sys{} detected all violations with few inputs (i.e., less than a second), illustrating the importance of further research on targeted test case generation.


\begin{table}[tbp]
    \small
    \setlength{\tabcolsep}{3.6pt}
    \center
    \begin{tabular}{c|c|c|c|c|c|c|c}
        Violation & V1                & V1.1                 & V2                & V4                & V5-ret                            & MDS-LFB                & MDS-SB         \\
        Type      & \cite{Kocher2018} & \cite{Kiriansky2018} & \cite{Kocher2018} & \cite{Google2018} & \cite{Maisuradze2018,Koruyeh2018} & \cite{RIDL,ZombieLoad} & \cite{Fallout} \\
        \hline
        \hline
        \# Inputs & 6                 & 6                    & 4                 & 62                & 2                                 & 2                      & 12             \\
    \end{tabular}
    \caption{Detection of known vulnerabilities on manually-written test cases.
    \emph{\# Inputs} is the average minimal number of random inputs necessary to surface a violation.}
    \label{tab:known-vulns}
\end{table}

\subsection{Contract Sensitivity}
\label{subsec:eval-arch}


\begin{figure}[t]
    \begin{minipage}[t]{0.49\columnwidth}
        \begin{lstlisting}[]
a = array1[b]
if (...)
  c = array2[a]\end{lstlisting}
        \subcaption{\ctseq{} violation}
    \end{minipage}
    \begin{minipage}[t]{0.49\columnwidth}
        \begin{lstlisting}[numbers=none]
if (...)
  a = array1[b]
  c = array2[a]        \end{lstlisting}
        \subcaption{\archseq{} violation}
    \end{minipage}
    \caption{Subtle difference in sensitivity of different contracts.}
    \label{fig:arch-leakage}
\end{figure}

The classic Spectre V1 exploit~\cite{Kocher2018} relies on two speculative loads, where the address of the second leaks the value loaded by the first, as in~\figref{arch-leakage}b.
Hardware defenses based on speculative taint tracking (STT)~\cite{STT2019,Weisse2019} prevent such leaks, but they do not intend to prevent leaks of \emph{non}-speculatively loaded data, as in~\figref{arch-leakage}a.

\memseq{} and \ctseq{} contracts cannot be used to test STT-like defenses as they forbid speculative leakage of any information (i.e., both examples would violate them).
Instead, we implement \archseq{} (\secref{our-contracts}), which \emph{permits} exposure of non-speculative data, but \emph{forbids} leakage of speculatively loaded data.
When testing Skylake against \archseq{}, \sys{} indeed reports violations corresponding to the classic V1 gadget (\figref{arch-leakage}b) and does not report violations in~\figref{arch-leakage}a.


\section{Scope and Limitations}
\label{sec:discussion}

\myparagraph{False contract conformance (false negatives)}
In several tests, \sys{} did not detect violations (\tabref{main-fuzz}).
This {\em does not prove} the absence of leaks: it merely shows that the explored space contained no counterexamples.

We see two potential sources of false negatives when we expand to a broader range of testing targets:
(1) Noisy measurements: to detect Meltdown/Foreshadow, the executor will have to handle faults, which may pollute the microarchitectural state and make the hardware traces too noisy.
(2) Low frequency of counterexamples: some vulnerabilities require a complicated combination of events to observe the leakage (e.g., CrossTalk~\cite{Ragab2021}) or to trigger speculation (e.g., Floating Point Value Injection~\cite{ragab2021rage}).
Random sampling may be too slow to find a counterexample that would surface them.

A false negative is also possible when the leak is observable only through a certain side channel (other than L1D, used in this paper), and this channel is too noisy to produce stable hardware traces.
For example, it may be the case for port contention channels.
We note, however, that all known speculative vulnerabilities can be observed through multiple channels, hence L1D measurements are sufficient to detect them. 
Yet it is not guaranteed for the future, currently-unknown vulnerabilities.

\myparagraph{False contract violations (false positives)}
If the model incorrectly emulates ISA, it leads to false positives.
Due to this reason, we excluded from the tests some instructions that are not implemented correctly in Unicorn.
Non-determinism in the executor may also cause false positives. 
However, we inspected a few counterexamples in each of the experiments described in~\secref{evaluation} and found no false positives.

\myparagraph{Generation of effective inputs}
\sys{} applies several restrictions to improve input effectiveness (\chref{meth-eq-coverage}).
This limits the test diversity, and might cause false negatives.
To eliminate them, future work may develop a targeted generation method that ensures effectiveness via program analysis, similar to Spectector~\cite{Guarnieri2018} or Scam-V~\cite{Nemati2020a}.

\myparagraph{Pattern coverage}
We used hazards as a proxy for speculation.
Yet a hazard is not a sufficient condition for speculative leakage.
For example, to trigger Spectre V1, branch predictor must be mistrained, and the speculation must be long enough to leave a trace.
These preconditions are hard to control on commercial CPUs and, thus, high pattern coverage does not guarantee that speculation was exercised.
Improved heuristics to estimate the speculation opportunities in generated test cases might lead to better results.

\myparagraph{Other side-channels}
\sys{} currently supports only attacks on L1D caches.
For other side-channels, we have to implement them within the executor (e.g., execution port attacks require reading of the port load).
For certain speculative attacks, the executor would have to be modified (e.g., Meltdown requires handling of page faults).

\myparagraph{Scalability issues}
In future, adding more features to Revizor and expanding the range of testing targets may exacerbate the search complexity.
However, some optimizations may balance out the added overheads:
While covering more side channels will require more testing time, improving the input generation process will speed up the testing.
Moreover, tests in different adversarial scenarios can easily run in parallel, on different machines with the same CPU model.

\myparagraph{Granularity of measurements}
\sys{} currently collects hardware traces once, after the execution of a test case.
It means that \sys{} does not record the information that could be potentially exposed by the order of memory accesses.
To observe the order, we could have probed caches concurrently with the test case execution, but it would introduce additional noise.
Hence, we opted for probing the final cache state, which is more deterministic, but records less information.


\section{Related Work}
\label{sec:related}

\myparagraph{Black-box detection of microarchitectural leaks} 
Several tools test black-box CPUs to find speculative vulnerabilities:
Medusa~\cite{Moghimi2020a} is a fuzzer for detecting variants of MDS\@.
SpeechMiner~\cite{Xiao2020} is a tool to analyze speculative vulnerabilities.
Both of them target specific attacks, while \sys{} detects vulnerabilities as violations of a contract. 

ABSynthe~\cite{Gras2020} and Osiris~\cite{Weber2021} automatically discover unknown side channels. In contrast, \sys{} detects unknown speculative leaks into a known side-channel (e.g. L1D cache). 

Scam-V~\cite{Nemati2020a} is a tool for testing CPUs against a model of side-channel leakage. Their approach is similar to \method{}, but their leakage model does not encompass speculation and they focus on analyzing simple, in-order CPUs (Cortex-A53) in which they identify unexpected leaks~\cite{Nemati2020}. 

\myparagraph{White-box detection of microarchitectural leaks}
A number of approaches use white-box information to detect microarchitectural leaks.
Fadiheh et al.~\cite{Fadiheh2019} proposed a SAT-based bounded model checker to find covert channels in RTL designs (in our terminology, they check RTL against \archseq{}).
CheckMate~\cite{Trippel2018} searches for pre-defined vulnerability patterns in CPU designs.
These tools are not applicable to testing of commercial black-box CPUs.

\myparagraph{Detection of architectural vulnerabilities}
Several tools fuzz for architectural vulnerabilities and ISA violations:
TestRIG~\cite{Woodruff2018} performs random testing of RISC-V designs.
Coppelia~\cite{Zhang2018} generates software exploits for CPU designs.
RFuzz~\cite{Laeufer2018} is a tool for fuzzing on the RTL level.

Formal models of the ISA~\cite{Armstrong2019,Degenbaev2012,Goel2017} could be augmented to capture speculation contracts, along the lines of our instrumentation of Unicorn.

\myparagraph{Information-flow checking}
While information-flow checking (verification and testing) of individual programs is a well-established field (see, e.g.,~\cite{sabelfeld2003language,backes2009automatic,barthe2011secure}) information-flow checking of language runtimes or processors (which requires reasoning about {\em all} programs) has not been widely studied. Notable approaches are~\cite{hritcu2013testing}, which generate random programs to surface non-interference violations in an information-flow monitor, and~\cite{Zhang2015} who propose to add information flow annotations to Verilog to detect timing leaks at compile time.

\myparagraph{Speculative leaks in software}
Several tools target detection of speculative leaks in software~\cite{Wang2018,He2019,Guarnieri2018,Oleksenko2020,Cauligi2020,Vassena2020}. 
They all rely on (sometimes implicit) assumptions about the speculation in hardware, see~\cite{cauligi2021sok} for an overview. \sys{} gives a first principled foundation for validating such assumption on black-box CPUs.


\section{Conclusion}
\label{sec:conclusion}

We presented \methodlong{} (\method{}), a technique to detect violations of speculation contracts in black-box CPUs.
We implemented \method{} in a framework called \sys{}, and used it to test Intel CPUs against a wide range of contracts.

Our experiments show that \sys{} effectively finds contract violations without reporting false positives.
The detected violations include known vulnerabilities such Spectre, MDS, and LVI, as well as novel variants.
This demonstrates that \method{} is a promising approach for third-party assessment of microarchitectural security in black-box CPUs.

Our work opens several avenues for future research, such as white-box analysis of emerging CPUs and mechanisms for secure speculation, coverage, and targeted testing, for which the open-source release of \sys{} will provide a solid foundation.

\begin{acks}
    We would like to thank Caroline Trippel and the anonymous reviewers for the constructive feedback, and Amaury Chamayou, Sylvan Clebsch, Manuel Costa, C{\'e}dric Fournet, Marco Guarnieri, Nuno Lopes, Saidgani Musaev, Robert Norton-Wright, and Alex Shamis for discussions and encouragement.

    This work was funded in part by DFG grant 389792660 as part of TRR 248 (CPEC); the Cluster of Excellence EXC 2050/1 (CeTI, project ID 390696704, as part of Germany’s Excellence Strategy); the Cloud-KRITIS Project funded by the S\"achsische Aufbaubank. This work was also supported by the Technion Hiroshi Fujiwara Cyber Security Research Center and the Israel National Cyber Directorate. We gratefully acknowledge support from Israel Science Foundation (Grant 1027/18).
\end{acks}

    \appendix

\section{Artifact Appendix}
\label{sec:artifact-appendix}

\subsection{Abstract}

The artifact for this paper includes the source code of Revizor, a set of scripts for reproducing the results, and a description of how to use them.
They help to reproduce the contract violations described in the paper and validate the claimed fuzzing speed.
\subsecref{artifact-new-variant} additionally describes a new violation discovered during the artifact evaluation.

\subsection{Artifact Meta-Information}
\label{subsec:artifact-check-list}

{\small
    \begin{itemize}
        \item {\bf Algorithm:} Random testing of CPUs
        \item {\bf Hardware:} x86 Intel CPU
        \item {\bf Metrics:} Detected contract violations and testing speed
        \item {\bf Output:} The test results and the violating test cases
        \item {\bf How much disk space required?:} less than 1GB
        \item {\bf How much time is needed to prepare workflow (approximately)?:} 1 hour
        \item {\bf How much time is needed to complete experiments (approximately)?:} 10 days
        \item {\bf Publicly available?:} Yes
        \item {\bf Code licenses (if publicly available)?:} MIT
        \item {\bf Workflow framework used?:} Yes
        \item {\bf Archived (provide DOI)?:} \code{10.5281/zenodo.5865606}
    \end{itemize}
}


\subsection{Description}\label{subsec:description}

Below is a brief description of the artifact.
You can find more details in the artifact's README file.

\myparagraphnodot{How to access:}

\noindent
\url{github.com/hw-sw-contracts/revizor-artifact}

\myparagraph{Hardware dependencies}
The artifact requires at least one physical machine with an Intel CPU and with root access.
Preferably, there should be two machines, one with an 8th generation (or earlier) Intel CPU and another with a 9th gen (or later) Intel CPU.
To have stable results, the machine(s) should not be actively used by any other software.

\myparagraphnodot{Software dependencies:}
\begin{itemize}
    \item Linux v5.6+ and Kernel Headers
    \item Unicorn Engine 1.0.2+ with Python bindings
    \item Python 3.7+ with packages \code{pyyaml,} \code{types-pyyaml,} \code{numpy,} \code{iced-x86, mypy}
    \item Bash Automated Testing System
\end{itemize}

\myparagraphnodot{System Configuration (Optional):}
For more stable results, disable hyperthreading and boot the kernel on a single core.

\subsection{Installation}\label{subsec:installation}

\begin{enumerate}
\item Get submodules:
\begin{verbatim}
# from the project's root directory 
> git submodule update --init --recursive
\end{verbatim}

\item Copy the ISA description:
\begin{verbatim}
> cp revizor/src/executor/x86/base.xml 
revizor/src/instruction_sets/x86
> cp revizor/src/executor/x86/base.xml 
x86.xml\end{verbatim}

\item Install the executor:
\begin{verbatim}
> cd revizor/src/executor/x86
> sudo rmmod x86-executor  
> make clean && make
> sudo insmod x86-executor.ko\end{verbatim}

\end{enumerate}

\subsection{Evaluation and Expected Results}\label{subsec:evaluation-and-expected-results}

The results of all next experiments will be stored in a corresponding subdirectory of \code{results/} with a timestamp.
For example, if you run Experiment 1 on 01.01.2022 at 13:00, the result will be stored in:
\code{results/experiment\_1/22-01-01-13-00}

This directory will contain the experiment logs, detected violations, and aggregated results (when applicable).

\subsubsection{Reproducing fuzzing results}

The following script will test each of the target-contract combinations in Table 3:
\begin{verbatim}
./experiment_1_main/run.sh
\end{verbatim}

Note that the last target (here called \code{target7-8}) is dependent on the machine.
If you execute the script on an 8th gen (or earlier) CPU, it will correspond to Target 7 in the table.
Otherwise, it will correspond to Target 8.

{\bf Note}: The violations of Targets 3 and 6 (called V1-var and V4-var in the paper) are very rare, and there is only a low chance that you will be able to reproduce them.
Unfortunately, such unpredictability of the results is an unavoidable consequence of random testing.

\subsubsection{Reproducing speculative store eviction} 

For this experiment, you will need a 9th gen Intel CPU or later (in the paper, we tested i7-9700).

To reproduce the violation reported in~\subsecref{validating-assumptions-about-speculative-execution}, execute the following script, which will test the CPU against a version of CT-COND that does not permit cache eviction by speculative stores.

\begin{verbatim}
./experiment_2_speculative_store_eviction/run.sh
\end{verbatim}

The expected result is that the execution detects a violation within an hour.
(If you run this command on an earlier Intel CPU, the expected result is no violations.)

\subsubsection{Fuzzing speed and detection time}

\begin{enumerate}
\item To measure the fuzzing speed, simply run Revizor for an hour in a configuration that does not find violations:
\begin{verbatim}
> ./revizor/src/cli.py fuzz -s x86.xml 
-i 200 -n 100000 --timeout 3600 
-v -c test-nondetection.yaml
\end{verbatim}

Upon completion, \sys{} will report the number of executed test cases and the number of inputs.

\item To measure the detection speed, execute:
\begin{verbatim}
./experiment_3a_detection_speed/run.sh
\end{verbatim}

The results are expected to approximately match Table 4.
The reported numbers are the mean values of the amount of time to detect each of the violations.
The meaning of the rows called "mds-*" depends on the target machine:
If the experiment is executed on an 8th gen (or earlier) CPU, they represent MDS-type vulnerabilities.
Otherwise, they represent LVI-type.

\item To measure the detection speed on handwritten test cases, execute:
\begin{verbatim}
./experiment_3b_handwritten_test_cases/run.sh
\end{verbatim}

The results are expected to approximately match Table 5.
The reported numbers are the average, median, minimum, and maximum number of inputs that was required to detect each of the violations with the given test case.
The exact numbers will differ slightly with each execution of this experiment, because the input generation seeds are generated randomly.

Note: The last two test cases (MDS-SB and MDS-LFB) work only on an 8th gen (or earlier) Intel CPU, because the later generations are patched against MDS.

\end{enumerate}

\subsubsection{Reproducing \archseq{} violations}

To test the CPU against \ctseq{} and \archseq{}, execute:
\begin{verbatim}
./experiment_4_arch_vs_ct/run.sh
\end{verbatim}

The expected result is that both contracts are violated.
You can find the counterexamples for both contracts in the results' directory, named \code{ct-seq-violation.asm} and \code{arch-seq-violation.asm}.

\subsection{Novel Variant of Store Bypass}
\label{subsec:artifact-new-variant}

When an anonymous reviewer evaluated the first experiment, they encountered a violation of \ctsbp{} by Target 2, which we did not observe in our previous experiments.
Under investigation, it appeared to be a new variant of Speculative Store Bypass.
The existing microcode patch provided by Intel mitigates this variant.

Below is a pseudocode of this violation:

\begin{verbatim}
1. *addr_slow = new_value;
2. x1 = *addr_fast;             
3. x2 = *addr_slow;  
4. y = array[x1 - x2];  
\end{verbatim}

Here, \code{addr\_fast} is a pointer to an address in memory initialized with \code{old\_value}.
When this code is executed, the pointer is already assigned with the address. 
\code{addr\_slow} is another pointer assigned with the same address.
However the address is calculated dynamically and hence takes more time to resolve.

At line~1, the store overrides the memory value with \code{new\_value}.
As the address take a long time to calculate, this store is delayed.

At line~2, the load fetches from the address.
The CPU may make a prediction that \code{addr\_fast} and \code{addr\_slow} do not alias, and proceed to speculatively fetch the now-outdated \code{old\_value};
this is the original Speculative Store Bypass.

At line~3, the CPU detects that lines~1 and 3 use the same address, and forwards the \code{new\_value} directly, without waiting for the address to resolve.

As a result, two consecutive loads from the same address speculatively return two different values.
The difference between the values is exposed into a side channel by line~4.

\sys{} labeled it as a violation of \ctsbp{} because the leakage is only possible when one of the loads bypasses the store, but not both of them.
This constitutes a violation of \ctsbp{}, where \emph{all} loads can bypass an aliasing store.

    \bibliographystyle{ACM-Reference-Format}
    \balance 
    \bibliography{ms}

\end{document}